\documentclass[5p,final,english]{elsarticle}

\usepackage[T1]{fontenc}
\usepackage[latin9]{inputenc}
\usepackage{graphicx}
\usepackage{verbatim}
\usepackage{amsmath}
\usepackage{amssymb}
\usepackage{amsfonts}
\usepackage{mathrsfs}
\usepackage{amsmath} 
\usepackage{color}
\usepackage{bm} 
\usepackage{amssymb}
\usepackage{xspace}
\usepackage{xcolor}
\usepackage{eucal}
\usepackage{hyperref}
\usepackage{babel}
\usepackage{float}
\usepackage{bbold}
\usepackage{tabu}
\usepackage{textcomp}
\usepackage{dcolumn}

\usepackage[caption=false]{subfig}

\begin{document}



\title{Singlet superconductivity enhanced by charge order in nested twisted bilayer graphene Fermi surfaces}

\author[add1,add2]{Evan Laksono}

\author[add1,add2]{Jia Ning Leaw}

\author[add3]{Alexander Reaves}

\author[add3]{Manraaj Singh}

\author[add1,add2]{Xinyun Wang}

\author[add1,add2,add3]{Shaffique Adam}

\author[add1,add2]{Xingyu Gu}

\address[add1]{Centre for Advanced 2D Materials and Graphene Research Centre, National University of Singapore, 117546, Singapore}
\address[add2]{Department of Physics, National University of Singapore, 117542, Singapore}
\address[add3]{Yale-NUS College, 16 College Avenue West, 138527, Singapore}

\begin{abstract}
Using the continuum model for low energy non-interacting electronic structure of moir\'e van der Waals heterostructures developed by Bistritzer and MacDonald~\cite{bistritzer2011moire} , we study the competition between spin, charge, and superconducting order in twisted bilayer graphene.  Surprisingly, we find that for a range of small angles inclusive of the so-called magic angle, this model features robust Fermi pockets that preclude any Mott insulating phase at weak coupling.  However, a Fermi surface reconstruction at $\theta \gtrsim 1.2^{\circ}$ gives emergent van Hove singularities without any Fermi pockets.  Using a hot-spot model for Fermi surface patches around these emergent saddle points, we develop a random-phase approximation from which we obtain a phase diagram very similar to that obtained recently by Isobe, Yuan, and Fu using the parquet renormalization group~\cite{isobe2018superconductivity} but with additional insights.  For example, our model shows strong nesting around time-reversal symmetric points at a moderate doping of $\sim 2 \times 10^{11}~{\rm cm}^{-2}$ away from the van Hove singularity.  When this nesting dominates, we predict that charge-order enhances singlet superconductivity, while spin-order suppresses superconductivity. Our theory also provides additional possibilities for the case of unnested Fermi surfaces.

\end{abstract}

\maketitle


\section{Introduction}
\vspace{-5pt}

Recently, superconductivity (SC) and correlated insulating phases have been reported in twisted bilayer graphene (tBG) at small twist angles~\cite{cao2018correlated, cao2018unconventional}.  The insulating gap was observed when the first electron moir\'e band is half-filled with two electrons (or holes) per moir\'e unit cell.  This is unexpected from the non-interacting picture, where the insulating phase occurs only when the first moir\'e band is completely filled and arises from the avoided crossing between the first and second moir\'e bands~\cite{wong2015local}.  The superconductivity was observed at slight doping ($~\sim 10^{11}\text{ cm}^{-2}$) away from the half-filled moir\'e band.  The strong resemblance between the observed phase diagram in tBG and that of high-$T_{c}$ cuprate and pnictide superconductors raises the tantalizing possibility that they share the same underlying mechanism.  If true, then the relative theoretical simplicity of tBG (nothing more than two rotated sheets of carbon), as well as the relative ease to prepare and manipulate experimental samples, makes this an unprecedented platform to probe and understand high-$T_{c}$ superconductivity.  In recent weeks, there have been a flurry of theoretical efforts that attempt to understand the correlated phenomenon in this system including Refs.~\cite{xu2018topological,dodaro2018phases,guo2018pairing,xu2018kekul,yuan2018model,po2018origin,koshino2018maximally,Volovik2018,baskaran2018theory,isobe2018superconductivity,you2018superconductivity,ray2018wannier,wu2018theory,liu2018chiral,thomson2018triangular,gonzalez2018kohn}.  However, it is fair to say that the  nature of the insulating phase, and the symmetry and mechanism of the superconducting pairing is still very much under debate.

The emergence of SC and insulating phases can be understood either from a strong coupling picture or at weak coupling.  At strong coupling, an antiferromagnetic Mott insulating phase arises for Dirac fermions at charge neutrality from the short-range part of the Coulomb interaction (which is enhanced in tBG compared to monolayer graphene)~\cite{tang2018the}.  Away from charge neutrality and at the half-filled moir\'e band, the SC can be understood as an emergent anti-ferromagnetic Heisenberg model where the pairing mechanism in the SC channel appears within mean field theory through a Fierz identity~\cite{xu2018topological}. In the weak coupling theories (see e.g. Refs.~{\cite{isobe2018superconductivity,ray2018wannier,liu2018chiral}), the instability is due to the saddle points and near nesting of the non-interacting Fermi surface (FS). The strong particle-hole (p-h) fluctuation resulting from saddle points and FS nesting leads to density wave instabilities, which are predicted to be responsible for the insulating phase observed in experiment. The p-h fluctuations can also pair two electrons together to form SC.   Since the non-interacting band-structure is known (see e.g. Ref.~\cite{bistritzer2011moire} and references therein), we focus here on weak coupling, while cognizant that a strong coupling mechanism could be operational in the experiments.   

In this work, we study the Fermi surface topology using the Bistritzer and MacDonald continuum low energy model. At $\theta=1.05^{\circ}$, we find robust non-nested Fermi pockets around Brillouin zone center when the chemical potential crosses saddle points.  Based on this finding, we predict that no insulating phase should be observed at magic angle $\theta=1.05^{\circ}$ in the weak coupling limit.  To study the insulating and superconducting phases, we construct a hot-spot RPA model.  If the Fermi surface is nested -- as is widely believed -- we predict that the insulating phase is charge-ordered and the superconductivity is a spin-singlet.  Our model also allows us to explore other possibilities when the Fermi surface is not nested.

\noindent


\begin{figure}
\begin{centering}
\subfloat[]{\label{fig:plotFS105}\includegraphics[width=0.22\textwidth]{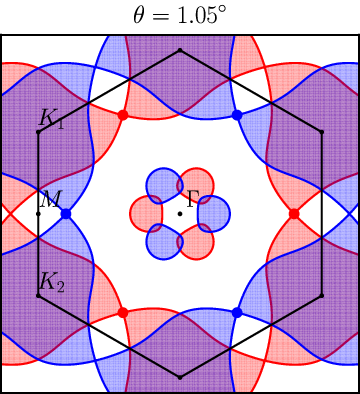}

}\subfloat[]{\label{fig:plotFS135}\includegraphics[width=0.22\textwidth]{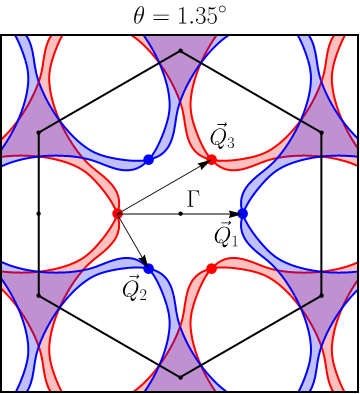}

}\\
\subfloat[]{\label{fig:ExchangeTerms}\includegraphics[width=0.22\textwidth]{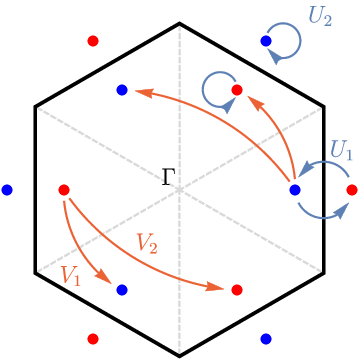}

}
\par\end{centering}
\caption{\textbf{Fermi surface reconstruction as a function of angle for twisted bilayer graphene}.  (a) The Fermi surface of the so-called ``magic angle" of $\theta=1.05^{\circ}$ when doped to the Van Hove energy.  There are robust Fermi pockets close to the moir\'e Brillouin zone center. No nesting is observed for these Fermi pockets, and they are expected to remain gapless at weak coupling. Accordingly, no Mott insulating phase should be observed at this twist angle and chemical potential.  (b)  Fermi surface at    
twist angle $\theta=1.35^{\circ}$ also at the Van Hove energy.  At this angle the Fermi surface has a different topology with the Fermi pockets having merged with the saddle points.  In this case, the weak coupling theory for twisted bilayer graphene is determined solely by the intra- and inter-saddle point couplings shown in (c).  Here $U_{1}$ is the exchange interaction between the saddle point and its time-reversal partner, $U_{2}$ is the density-density interaction. $V_{1}$ and $V_{2}$ are the exchange interactions between two saddle points separated by $\vec{q}=\vec{Q}_{2}$ and $\vec{Q}_{3}$ marked in panel (b).\label{fig:Fig1}}
\end{figure}

\begin{figure}
\begin{centering}
\subfloat[]{\includegraphics[width=0.22\textwidth]{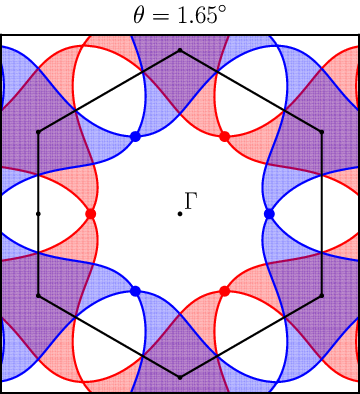}

}\subfloat[]{\label{fig:Nesting}\includegraphics[width=0.22\textwidth]{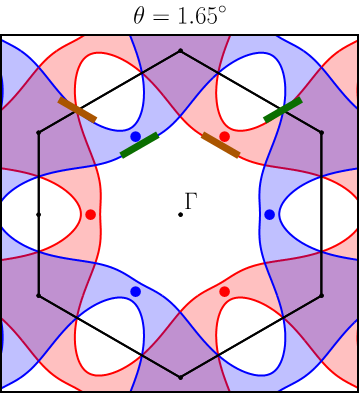}

}
\par\end{centering}
\begin{centering}
\subfloat[]{\includegraphics[width=0.22\textwidth]{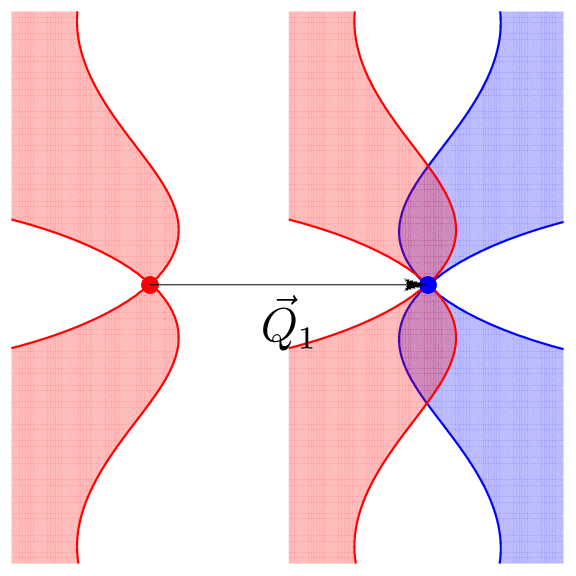}

}\subfloat[]{\includegraphics[width=0.22\textwidth]{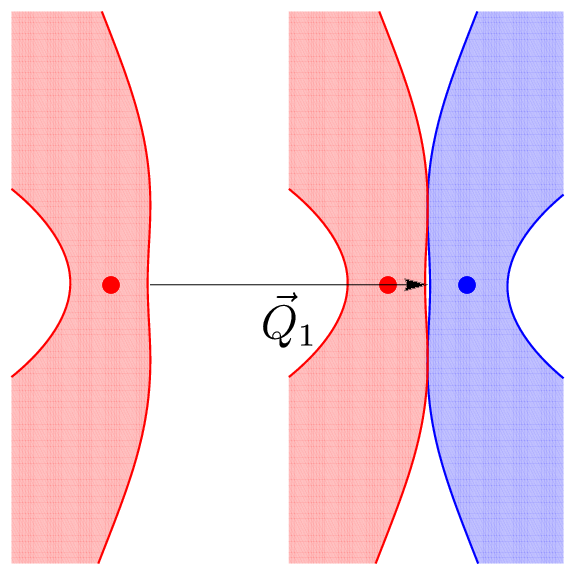}

}
\par\end{centering}
\caption{\textbf{Strong nesting for Fermi surface at a chemical potential slightly away from the saddle points.} (a) The Fermi surface for twist angle $\theta=1.65^{\circ}$ with the doping at the Van Hove energy.  (b) Modified Fermi surface after a slight doping of $n=2\times10^{11}\:\text{cm}^{-2}$ away from the saddle point. (c) Zoom-in of (a) showing weak $\vec{Q}_{1}$ nesting at precisely the Van Hove energy.  (d) Zoom-in of (b) showing strong nesting of the flat regions.  Doping away from the Van Hove energy is necessary to enhance the $\vec{Q}_{1}$ nesting which is widely believed to be responsible for the Mott insulating and superconducting phases. \label{fig:FS165}}
\end{figure}

\section{Model}
\label{section::model}
\vspace{-10pt}
\subsection*{Fermi surface of non-interacting bands}
\vspace{-5pt}

Interlayer coupling between TBG layers results in a hybridization between the Dirac bands which belong to the constituent layer. The essential features of the hybridized bands can be captured by three Fourier components of similar magnitude $w$, consistent with the underlying three-fold rotational symmetry of the tBG~\cite{bistritzer2011moire, santos2007graphene}. Our non-interacting model of tBG is fully based on the continuum model of Ref. \cite{bistritzer2011moire}, in which the Hamiltonian takes the following form:

\begin{equation}
H_{s}(\vec{k}_{1},\vec{k}_{2};\theta) =
\begin{pmatrix}
h_{s}(\vec{k}_{1};\frac{\theta}{2}) & T_{s}(\vec{r})\\
T_{s}^{\dagger}(\vec{r}) & h_{s}(\vec{k}_{2};-\frac{\theta}{2})\\
\end{pmatrix},
\end{equation}

\noindent where $\vec{k}_{1}$ ($\vec{k}_{2}$) refers to the momentum states of the first (second) layer, while $s = \pm$ denotes the valley K/K'. We use the following convention in our work:

\begin{subequations}
\begin{align}
h_{s}(\vec{k},\theta) = s v_F k
\begin{pmatrix}
0 & e^{-is(\theta + \theta_{\vec{k}})}\\
e^{is(\theta + \theta_{\vec{k}})} & 0,
\end{pmatrix},\\
T_{s}(\vec{r}) =  w \sum_{n}\exp(-is\vec{q}_{n}\cdot\vec{r})
\begin{pmatrix}
e^{i\frac{2ns\pi}{3}} & 1\\
e^{-i\frac{2ns\pi}{3}} & e^{i\frac{2ns\pi}{3}}\\
\end{pmatrix},
\end{align}
\end{subequations}

\noindent where $v_{F}$ is the Fermi velocity of monolayer graphene, and $n \in \{0,1,2\}$. The vectors  $\vec{q}_{n} = 2K\sin(\frac{\theta}{2})[\mathcal{R}(\frac{2n\pi}{3})\{0,1\}]$ define the connections between momenta on different layers, i.e. $\vec{k}_{2} = \vec{k}_{1}+ s\vec{q}_{n}$ and $\mathcal{R}(\phi)$ is the rotation operator.  In our calculations, for the single adjustable parameter, we use the accepted value of $w = 110$ meV.

Under the influence of interlayer couplings, the bands experience significant reconstruction, such as avoided crossings at momentum states located midway between the Dirac points, Van Hove singularities (VHS) in the low-energy spectrum and Fermi velocity renormalization (see e.g. Ref.~\cite{wong2015local} and references therein).  Interestingly, the hybridization can lead to a total suppression of the kinetic energy of the low-energy electrons at some so-called ``magic angles".  Although this is widely believed to be germane to the experiments, our results show that there is nothing special about the magic angles. In fact, within weak coupling, (as we explain below) we expect no insulating phase to occur at the magic angles. 


The hybridized bands show a complex structure.  When doped away from the charge neutrality point, Fermi pockets start forming around the Dirac points.  With further doping, these Fermi pockets become connected and saddle points emerge.  Due to the three-fold symmetry of tBG, these saddle points occur at three momenta connected by symmetry in the Brillouin zone for each valley. The doped regions and the saddle points are plotted in Fig.~(\ref{fig:Fig1}) as red for one valley and blue for the other valley.  These saddle points exhibit themselves as the VHS in the density of states (DOS) or peaks in the local DOS~\cite{wong2015local}.

For small angles $\theta\lesssim 1.07^\circ$, we show (for the first time) that the FS at this doping contains 6 small Fermi pockets around the $\Gamma$ point, in addition to those at the Dirac points.  In particular, these small Fermi pockets are present when the twist angle is at the magic angle $\theta=1.05^\circ$ (see Fig.~\ref{fig:plotFS105}).  Since there is no nesting that might gap out these Fermi pockets, within weak coupling, we expect no insulating phase at these small twist angles.  The FS contour undergoes complicated evolution for twist angle $1.07^\circ<\theta<1.15^\circ$.  For the range of twist angle $1.15^\circ < \theta < 1.34^\circ$ the small Fermi pockets disappear, and the FS contains 12 saddle points.  When the twist angle is further increased, we observe that the number of saddle points in the FS reduces from 12 to 6, {\it without} any small Fermi pockets, consistent with recent predictions by Ref.~\cite{gonzalez2018kohn}.  The FS at $\theta=1.35^{\circ}$ is shown in Fig.~(\ref{fig:plotFS135}).  For the rest of this paper we focus on this regime of 6 saddle points without the Fermi pockets, as this is the most generic situation that is favorable to the observation of correlated physics. 

Due to the logarithmically divergent DOS near the saddle points, the role of interaction is enhanced in this region. The p-h fluctuation is strong for momentum connecting these saddle points ($\vec{Q}_1$,$\vec{Q}_2$ and $\vec{Q}_3$ in Fig.~(\ref{fig:plotFS135})) because the susceptibility also diverges logarithmically at these momenta~\cite{rice1975new}. The fluctuation leads to instabilities such as charge density wave (CDW) and spin density wave (SDW).  The fluctuations are weakened upon further doping, resulting in the suppression of long-range charge (spin) order. However, the remaining short range fluctuations can lead to pairing between two electrons which gives rise to SC phase. Since the region of the FS with the largest DOS is expected to determine the correlated phase, as a first approximation, we construct a patch model around the saddle points.  The patch model provides insight about the pairing symmetry without considering the whole FS (see e.g. Ref.~\cite{maiti2013superconductivity} for application of the patch model to other systems).  

In addition to the appearance of saddle points, another feature of the FS is possible nesting. The nested FS can further enhance the p-h fluctuation.  Several authors including Refs.~\cite{ray2018wannier,liu2018chiral} assume that nesting is dominant, i.e. $\chi(\vec{Q}_1)\gg \chi(\vec{Q}_2),\chi(\vec{Q}_3)$. Here $\chi$ is the susceptibility, and the nesting vectors $\vec{Q}_1$, $\vec{Q}_2$ and $\vec{Q}_3$ are shown in Fig.~\ref{fig:FS165} for $\theta=1.65^\circ$.  We find $\vec{Q}_1$ nesting is poor when the energy crosses saddle points (see Fig~\ref{fig:FS165}c).  The nesting gets better when the energy is slightly away from the saddle points as shown in Fig~\ref{fig:FS165}d.
In the remaining part of this paper, we first study case where $\vec{Q}_1$ is the nesting vector, then study the case that FS is not nested. In this latter case, whether $\vec{Q}_1$, $\vec{Q}_2$ or $ \vec{Q}_3$ is dominant in p-h channel depends on the details of the system. We explore the consequence of each possibility in this paper.

\subsection*{Interaction model}
\vspace{-5pt}

In general, there are three standard methods to study p-h fluctuation mediated superconductivity: (1) spin-fermion model, (2) parquet renormalization group (p--RG) and
(3) the RPA (that we adopt here).  The second method was applied to tBG recently by Ref.~\cite{isobe2018superconductivity}.  We find that our RPA method (adapted from Refs.~\cite{graser2009s,kemper2010af,scalapino1986d} and applied to tBG) is qualitatively similar to the results from p-RG~\cite{isobe2018superconductivity} and consistent with the intuition from spin-fermion model.  However, it provides additional insights from which we can make predictions for experiments.

There are 9 independent intra-patch
and inter-patch interactions, as shown in Ref.~\cite{isobe2018superconductivity}. These interactions can lead to different
instabilities including charge (spin) density wave, superconductivity
and pair density wave (PDW). 
Using the linearized gap equation method~\cite{nandkishore2012chiral,ganesh2014theoretical}, the interaction strength in each instability channel is shown in Ref.~\cite{isobe2018superconductivity}.  Among all these instabilities, we don't consider PDW phase because we don't believe it is relevant for the experimental system. When $\vec{Q}_1$ is the nesting vector, among all the 9 interactions,
we only consider 2 non-zero interactions between time reversal patches: exchange interaction
$U_{1}$ and inter-patch density-density interaction $U_{2}$, which
correspond to $g_{41}$ and $g_{42}$ in Ref.~\cite{isobe2018superconductivity}. The interacting Hamiltonian is
\begin{align}
H_{int}^{\vec{Q}_1}=\frac{1}{2}\sum_{i=1}^{6}\sum_{\sigma\sigma^\prime}(U_1c^{\dagger}_{i,\sigma}c^{\dagger}_{\bar{i},\sigma^\prime}c_{i,\sigma^\prime}c_{\bar{i},\sigma}+U_2c^{\dagger}_{\bar{i},\sigma}c^{\dagger}_{i,\sigma^\prime}c_{i,\sigma^\prime}c_{\bar{i},\sigma}),
\end{align}

\noindent where $i$ is the patch index, $\bar{i}$ is the patch opposite to $i$, $\sigma$ and $\sigma^{\prime}$ are spin indexes.
The diagrammatic
representation of $U_{1}$ and $U_{2}$ is shown in Fig.~(\ref{fig:ExchangeTerms}).  These two interactions are non-vanishing because when $\vec{Q}_{1}$ is the dominant nesting vector, it is reasonable to only consider the density wave phase with momentum $\vec{Q}_{1}$, (called CDW$^\prime$ and SDW$^\prime$ in Ref.~ \cite{isobe2018superconductivity}). Comparing the interaction in superconductivity channel and interaction in density wave channel, $g_{41}$ and $g_{42}$ are the two terms shared in common and should dominate when we study the interplay between density wave phase and SC phase.  Using this simplification, the interaction strength in each instability channel becomes very simple: $V_{CDW}=4U_{1}-2U_{2}$,$V_{SDW}=2U_{2}$, $V_{singlet-SC}=2(U_{2}+U_{1})$, $V_{triplet-SC}=2(U_{2}-U_{1})$.  When the FS is not nested, it is possible that $\vec{Q}_2$ or $\vec{Q}_3$ is the momentum with the largest p-h fluctuation. In this case, for the same reason, we consider only two non-zero interactions $V_1$ and $V_2$. The corresponding interacting Hamiltonian is:
\begin{align}
\nonumber H_{int}^{Q_{2(3)}}=&\frac{1}{2}\sum_{i=1}^{6}\sum_{\sigma\sigma^\prime}(V_1c^{\dagger}_{\bar{i}+2,\sigma}c^{\dagger}_{i+2,\sigma^\prime}c_{i,\sigma^\prime}c_{\bar{i},\sigma}\\ 
&+V_2c^{\dagger}_{i+2,\sigma}c^{\dagger}_{\bar{i}+2,\sigma^\prime}c_{i,\sigma^\prime}c_{\bar{i},\sigma})
\end{align}
The diagrammatic representation of $V_1$ and $V_2$ are shown in Fig.~(\ref{fig:ExchangeTerms}).

\section{Results and Discussion}
\label{section::result}
\vspace{-5pt}

\begin{figure}
\begin{centering}
\subfloat[]{\label{fig:Q1dominant}\begin{centering}
\includegraphics[width=0.22\textwidth]{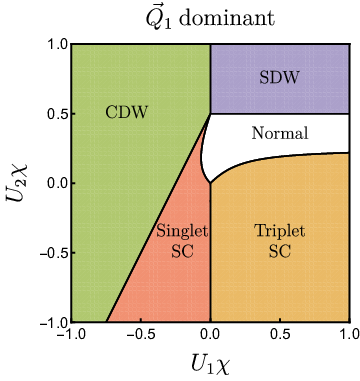}
\par\end{centering}
}\subfloat[]{\label{fig:phasediagram}\begin{centering}
\raisebox{0.4cm}{\includegraphics[width=0.2\textwidth]{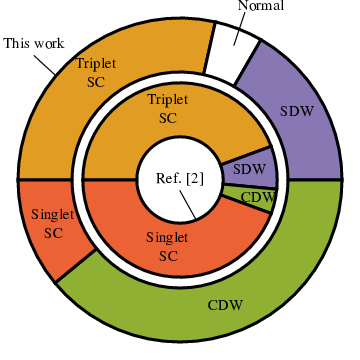}}
\par\end{centering}
}
\par\end{centering}
\begin{centering}
\subfloat[]{\label{fig:Q2dominant}\begin{centering}
\includegraphics[width=0.22\textwidth]{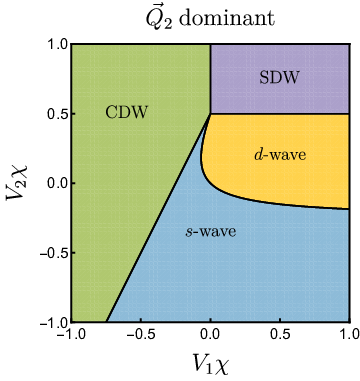}
\par\end{centering}
}\subfloat[]{\label{fig:Q3dominant}\begin{centering}
\includegraphics[width=0.22\textwidth]{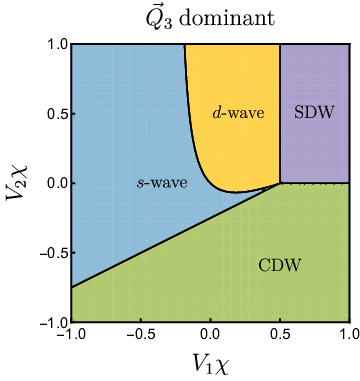}
\par\end{centering}
}
\par\end{centering}
\centering{}\caption{\textbf{Phase diagrams within the RPA-patch model for different dominant pairing momentum.} (a) The phase diagrams for the case where the susceptibility is peaked at pairing momentum $\vec{Q}_1$. Here, we find
that the charge-density-wave promotes the singlet superconducting
phase, while the spin-density-wave suppresses any superconducting
phase. (b) Outer circle shows a circular cut of (a) about the non-interacting point $U_1 = U_2 = 0$ (the angle represents the ratio between the exchange and density-density interactions).  Our phase diagram is similar to the predictions by Isobe et
al.~\cite{isobe2018superconductivity} using a parquet renormalization group method, which is plotted as the inner ring.  However, unlike Ref.~\cite{isobe2018superconductivity}, we find that the spin-density-wave and triplet superconductivity are separated by a normal phase, and this has strong implications for experiments.  In (c) and (d), we also show the RPA-patch model phase diagrams for when $\vec{Q}_2$ and $\vec{Q}_3$ are dominant, respectively. In both these cases, the degeneracy between the s-wave and d-wave are lifted, and these are enhanced by the charge-density-wave and spin-density-wave phases, respectively. The phase diagrams are slightly different from those in our published version because we missed a factor of $\frac{1}{2}$ there.}
\end{figure}

First, we consider the case where $\vec{Q}_1$ is the dominant nesting vector. 
The renormalized CDW and SDW susceptibility at RPA level is
\begin{subequations}
\begin{align}
\chi_{CDW}^{RPA}(\vec{Q}_1)=\frac{\chi_{ph}^{0}(\vec{Q}_1)}{1+V_{CDW}\chi_{ph}^{0}(\vec{Q}_1)}\\
\chi_{SDW}^{RPA}(\vec{Q}_1)=\frac{\chi^{0}_{ph}(\vec{Q}_1)}{1-V_{SDW}\chi^{0}_{ph}(\vec{Q}_1)}
\end{align}
\end{subequations}
where $V_{CDW}$ and $V_{SDW}$ are the interactions in CDW and SDW channel. $\chi_{ph}^{0}(\vec{Q}_1)=\sum_{i,j=1,2}\chi_{ph}^{ij}(\vec{Q}_1)$ is the bare static p-h susceptibility with $\chi_{ph}^{ij}(\vec{q})=\sum_{\vec{k}}
[f(\epsilon^{i}_{\vec{k}})-f(\epsilon^{j}_{\vec{k}+\vec{q}})]$ $/[\epsilon^{j}_{\vec{k}+\vec{q}}-\epsilon^{i}_{\vec{k}}]$.
$i,j=1,2$ are valley indices, while $f(\epsilon)$ is the Fermi-Dirac distribution function, with $\epsilon_{\vec{k}}^1$ and $\epsilon_{\vec{k}}^2$ are the dispersion for the two valleys.
Besides the nesting vector $\vec{Q}_1$, we observe other "accidental nested" segments contributing to large susceptibility, which is marked by thicker lines in Fig.~(\ref{fig:Nesting}). Since one of the two parallel segments is away from saddle points, we stick to saddle point physics in this paper and leave this nesting vector for future study. 
The renormalized SC susceptibility  from RPA like ladder diagrams is 
\begin{equation}
\chi_{i-SC}=\frac{\chi^{0}_{pp}(0)}{1+V^{irred}_{i-SC}\chi^{0}_{pp}(0)}
\end{equation}
where $i-SC$ represents different pairing symmetry. $V^{irred}_{i-SC}$ is the irreducible pairing vertex in each SC channel and the bare p-p susceptibility $\chi_{pp}^{0}(0)=\sum_{\vec{k}}[1-f(\epsilon_{\vec{k}})-f(\epsilon_{-\vec{k}})]/[\epsilon_{\vec{k}}+\epsilon_{-\vec{k}}]$.
$\chi^0_{pp}(0)$ is always logarithmically divergent as long as the system preserves time reversal symmetry.

The critical interaction strength of density wave is determined by $1\pm V_{CDW(SDW)}^{cri}\chi_{ph}^{0}(\vec{Q}_1)=0$. Due to the logarithm divergence of $\chi_{ph}^{0}$, this condition can be satisfied in the weak coupling limit. Once the interaction strength $V_{CDW(SDW)}$is larger than
the critical value, the system will be in the charge(spin) density
wave phase. We also notice for some \{$U_1$, $U_2$\}, interaction in CDW and SDW channel are both larger than the critical value. Since this case is outside the formal validity of the RPA theory, as is common, we take the one with the larger effective interaction to be the ground state because presumably its $T_c$ is higher. If the interaction strength is smaller than the critical
value, there will be no long range density wave order. However, the
short range charge (spin) density wave order can mediate the pairing
between two electrons. SC instability occurs when $\chi_{i-SC}$ diverges. Since $\chi_{pp}^{0}(0)$ is also logarithmically divergent, SC occurs as long as the irreducible pairing vertex is negative. To the first order, as we show above, $V^{irred}_{singlet-SC}=2(U_2+U_1)$ and $V^{irred}_{triplet-SC}=2(U_2-U_1)$. To get singlet SC, $U_1+U_2$ needs to be negative so strong electron phonon coupling is required. To get triplet SC, $U_1$ needs to be larger than $U_2$. This is impossible for finite range interactions because $U_1$ involves large momentum transfer. To get attractive interaction, we need to consider the renormalization of these irreducible pairing vertices. The renormalization to RPA level is 
\begin{subequations}
\begin{align}
& V_{singlet-SC}^{irred}=\frac{1}{2}V_{CDW}^{RPA}(\vec{Q}_1)+\frac{3}{2}V_{SDW}^{RPA}(\vec{Q}_1)\label{eq:RPA-Singlet-SC} \\
&V_{triplet-SC}^{irred}=-\frac{1}{2}V_{CDW}^{RPA}(\vec{Q}_1)+\frac{1}{2}V_{SDW}^{RPA}(\vec{Q}_1)\label{eq:RPA-Triplet-SC}
\end{align}
\end{subequations}
where $V_{CDW}^{RPA}=\frac{V_{CDW}}{1+V_{CDW}\chi^{0}_{ph}(\vec{Q}_1)}$ and
$V_{SDW}^{RPA}=\frac{V_{SDW}}{1-V_{SDW}\chi^{0}_{ph}(\vec{Q}_1)}$.
The phase diagram is shown in Fig.~(\ref{fig:Q1dominant}). We find singlet
SC is adjacent to the CDW while triplet SC and SDW are separated by normal state. This means CDW induces singlet superconductivity while SDW suppresses superconductivity. This can be  simply understood from Eq.(\ref{eq:RPA-Singlet-SC}) and Eq.(\ref{eq:RPA-Triplet-SC}): in the region close to CDW phase, CDW
fluctuation is large and the value of $V_{CDW}^{RPA}(\vec{Q}_1)$ is large and negative.
In this case, the interaction in singlet channel
becomes attractive while the interaction in triplet channel is repulsive.
On the other hand, in the region close to SDW, SDW fluctuation is
large and the value of $V_{SDW}^{RPA}(\vec{Q}_1)$ is large and positive. As a result,
both $V_{singlet-SC}^{RPA}$ and $V_{triplet-SC}^{RPA}$ are positive and SC is suppressed. This conclusion is consistent with Ref.~\cite{isobe2018superconductivity}, where CDW and singlet SC develop simultaneously while SDW and triplet develop independently. 

Our result is also consistent with the intuition
from the spin-fermion model. The spin fermion model states when the pairing is mediated by
spin fluctuation, the order parameter, as a function of momentum, will
change sign if the momentum is connected by the nesting vector. In
the other case, when the pairing is mediated by charge fluctuation,
the order parameter will not change sign. From the the form of order
parameter and pairing symmetry in \cite{isobe2018superconductivity}, we find there is no sign
change for all the singlet pairing, indicating they are mediated by charge fluctuation.  We note that our findings are also consistent with the experiments~\cite{cao2018correlated,cao2018unconventional}. Since the critical magnetic field is close to the Pauli limit, they believe that the SC phase is a spin singlet.  In this case, the charge ordered insulating 
phase makes tBG slightly different from cuprates and iron pnictides, where the insulating 
phases are magnetic ordered states.
Scanning probe microscopy experiments could confirm our expectation of a CDW insulating phase.

We also compare our result that obtained previously using p-RG. As shown in Fig.~(\ref{fig:phasediagram}), if we start
from the negative y axes and go counter-clockwise, the system will
be in triplet SC, SDW, CDW and singlet phase in sequence. Correspondingly, in the RG phase diagram,
if we go along the $d_{2-}=1.0$ line  from
$\pi$ to $-\pi$, these four phases appear in the same order.  The main difference is the normal phase seen in the RPA. 

\begin{figure}
\begin{centering}
\subfloat[]{\label{fig:intrasusceptibility165}\includegraphics[width=0.22\textwidth]{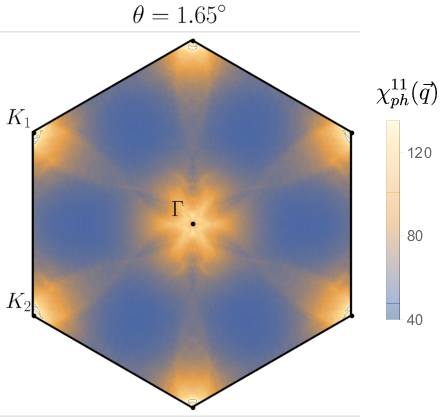}

}\subfloat[]{\label{fig:intersusceptibility165}\includegraphics[width=0.22\textwidth]{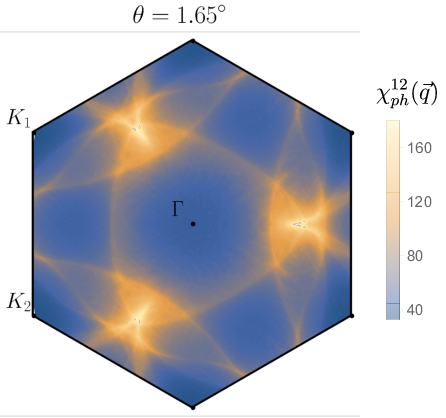}

}
\par\end{centering}
\begin{centering}
\subfloat[]{\label{fig:intrasusceptibility116}\includegraphics[width=0.22\textwidth]{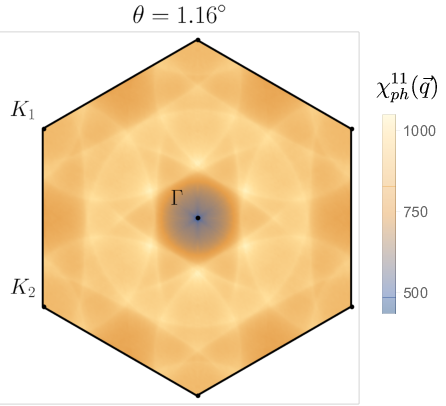}

}\subfloat[]{\label{fig:intersusceptibility116}\includegraphics[width=0.22\textwidth]{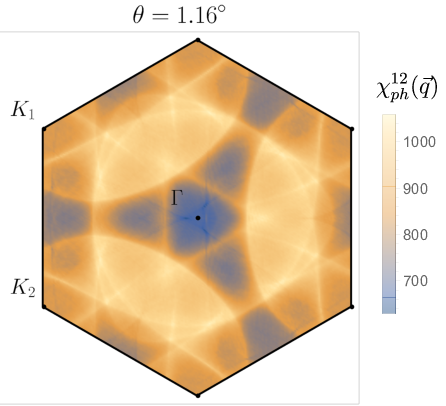}

}
\par\end{centering}
\caption{\textbf{The momentum dependence of the bare susceptibility.}  For twist angle $\theta = 1.65^{\circ}$ and doping at the Van Hove energy, we show the (a) intra-valley and (b) inter-valley bare susceptibility for the moir\'e band theory of Ref.~\cite{bistritzer2011moire}.  The bright spots in (a) close to $K$ points correspond to the momenta $\vec{Q}_3$, while
the bright spots in the inter-valley susceptibility corresponds to $\vec{Q}_1$ and $\vec{Q}_2$. Our patch model is built assuming that $\vec{Q}_1$, $\vec{Q}_2$ or $\vec{Q}_3$ is dominant.  At generic angles and doping, the susceptibility is peaked at other vectors. As a representative, intra-valley and inter-valley bare susceptibility for twist angle $\theta=1.16^\circ$ at half-filled moir\'e band are plotted in (c) and (d). These will be explored in a future work.}
\end{figure}

\begin{figure}
\begin{centering}
\subfloat[]{
\includegraphics[width=0.22\textwidth]{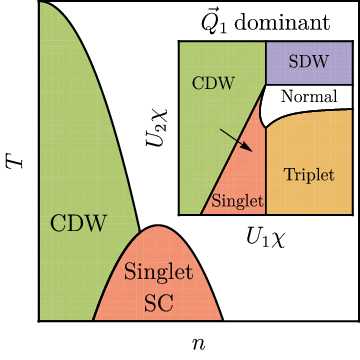}
}
\par\end{centering}
\begin{centering}
\subfloat[]{
\includegraphics[width=0.22\textwidth]{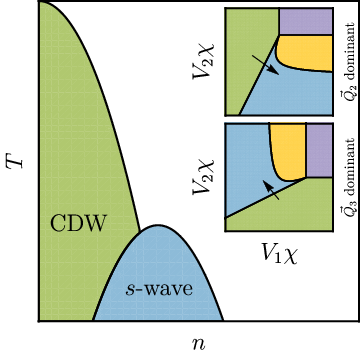}
} \subfloat[]{\begin{centering}
\includegraphics[width=0.22\textwidth]{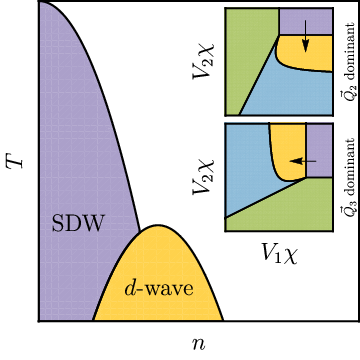}
\par\end{centering}
}
\par\end{centering}
\caption{\textbf{Three possibilities for saddle-point induced superconductivity and insulating phase in twisted bilayer graphene}. (a) shows the phase diagram for the case where the
susceptibility is peaked at $\vec{Q}_1$. In this case, the charge density wave facilitates the singlet pairing, while the spin density wave suppresses the superconducting phase.  The superconducting dome has singlet pairing, and the insulating phase is a charge density wave (with s-wave and d-wave being degenerate). The arrow shows the transition
direction when the doping density is increased. (b) and (c) show the possible phase diagrams in the case where the susceptibility is peaked at either $\vec{Q}_2$ or $\vec{Q}_{3}$. In both these cases, the insulating-superconducting transition could either be charge-density-wave and s-wave, or spin-density-wave and d-wave.}
\label{fig:ExperimentPD}
\end{figure}

Up to this point, we have discussed the case where $\vec{Q}_1$ is the nesting vector (as is widely believed).  We now consider the possibility of no nesting using the example of $\theta=1.65^{\circ}$ at the VHE as an example. We plot the distribution of $\chi^{11}_{ph}(\vec{q})$ and $\chi^{12}_{ph}(\vec{Q})$ for $\theta=1.65$ in Fig.~(\ref{fig:intrasusceptibility165}) and (\ref{fig:intersusceptibility165}) respectively, and that of $\theta=1.16$ in Fig.~(\ref{fig:intrasusceptibility116}) and (\ref{fig:intersusceptibility116}) for comparison. Without layer hybridization, the contribution from the inter-valley susceptibility is guaranteed to vanish for $\vec{q} = 2\vec{K}$, where $\vec{K}$ denotes the moir\'e Brillouin zone corner; however, in Fig. (4), we show that for Hamiltonian (1), the inter-valley susceptibility derived from a multi-orbital approach \cite{wu2015triplet} is computed to be finite for $\vec{Q}_1$ and $\vec{Q}_2$ (where both are away from 2$\vec{K}$), in agreement with the estimates of Ref. \cite{isobe2018superconductivity}.For $\theta=1.65$, there are clear peaks in the plot of $\chi^{11}_{ph}(\vec{q})$ at $\vec{Q}_3$ and the momenta related to $\vec{Q}_3$ by symmetry. The other peak at $\Gamma$ reflects the large DOS.  The large $\chi(0)$ will lead to ferromagnetic phase and induce triplet pairing~\cite{wu2015triplet}. We do not consider this case in our paper. For $\chi^{12}_{ph}(\vec{Q})$, the peaks corresponding to $\vec{Q}_1$ and $\vec{Q}_2$ are close to each other such that they merge into a single peak. Since $\chi(\vec{Q}_1)$, $\chi(\vec{Q}_2)$ and $\chi(\vec{Q}_3)$ are of the same order of magnitude, the dominant one would depend on the details of the system.  We have already discussed the case when $\vec{Q}_1$ is dominant.  For either $\vec{Q}_2$ or $\vec{Q}_3$   
dominant, the calculation is almost identical to that of $\vec{Q}_1$ except we replace $U_1$ and $U_2$ with $V_1$ and $V_2$.  For the SC part, if we focus on the singlet pairing claimed in the experiment, the irreducible pairing vertices for $\vec{Q}_2$ are
\begin{subequations}
\begin{align}
&V_{s-wave}^{irred}=\frac{1}{2}V_{CDW}^{RPA}(\vec{Q}_2)+\frac{3}{2}V_{SDW}^{RPA}(\vec{Q}_2)\label{eq:RPA-s-SC}\\
&V_{d-wave}^{irred}=-\frac{1}{2}V_{CDW}^{RPA}(\vec{Q}_2)-\frac{3}{2}V_{SDW}^{RPA}(\vec{Q}_2).\label{eq:RPA-d-SC}
\end{align}
\end{subequations}
(For $\vec{Q}_3$, we exchange $V_1$ and $V_2$).  The resulting phase diagrams are shown in Fig.~(\ref{fig:Q2dominant}) and (\ref{fig:Q3dominant}). In both cases, CDW induces $s$-wave pairing and SDW induces $d$-wave pairing. The two $d$-wave pairings $d_{xy}$ and $d_{x^{2}-y^{2}}$ are degenerate because they are the basis of a two dimensional representation of $D_{3}$, the point group for tBG~\cite{liu2018chiral,isobe2018superconductivity}. Considering that both $d_{xy}$ and $d_{x^{2}-y^{2}}$ have nodal lines in the spectrum, the true ground state should be $d_{xy}\pm id_{x^{2}-y^{2}}$, which is fully gapped and lower in free energy. 

In summary, assuming that correlated phenomenon in tBG is due to the saddle points in momentum space, we show that the magic angle is not germane. Fermi pockets at small angles including all the magic angles would suppress the insulating phase.  Our RPA-patch model predicts several possible combinations of insulating and superconducting phases shown in Fig.~\ref{fig:ExperimentPD}: (a) CDW with momentum $\vec{Q}_1$ and singlet pairing SC, (b) CDW with momentum $Q_{2(3)}$ and \textit{s}-wave SC, and (c) SDW with momentum $\vec{Q}_{2(3)}$ and \textit{d+id}-wave SC, depending on the details of the FS.  Generically, we expect that the $s$-wave pairing requires either $U_1$ or $V_1$ to be attractive. This is consistent with the intuition that $s$-wave pairing is suppressed by on-site repulsion thus making extra attraction (i.e. electron phonon coupling) necessary. Isotopic effect should be observed in the latter case. 
 
\section{Acknowledgements}
\noindent The authors thank Greg Stewart for discussions. 
This work is supported by the Singapore Ministry of Education MOE2017-T2-1-130 and MOE2017-T2-2-140.


\bibliographystyle{model1a-num-names}
\bibliography{ref_new}

\end{document}